\documentstyle[multicol,aps,prl]{revtex}
\begin{document}

\newcommand{\beq}{\begin{eqnarray}}
\newcommand{\eeq}{\end{eqnarray}}
\newcommand{\eqna}{\begin{eqnarray}}
\newcommand{\eqne}{\end{eqnarray}}
\newcommand{\dia}{\begin{displaymath}}
\newcommand{\die}{\end{displaymath}}
\newcommand{\eqnaa}{\begin{eqnarray*}}
\newcommand{\eqnae}{\end{eqnarray*}}
\def\dleft{\rlap{{\it D}}\raise 8pt\hbox{$\scriptscriptstyle\Leftarrow$}}
\def\prop{\propto}
\def\dright{\rlap{{\itD}}
\raise 8pt\hbox{$\scriptscriptstyle\Rightarrow$}}
\def\lrartop#1{#1\llap{
\raise 8pt\hbox{$\scriptscriptstyle\leftrightarrow$}}}
\def\_#1{_{\scriptscriptstyle #1}}
\def\&#1{^{\scriptscriptstyle #1}}
\def\sss{\scriptscriptstyle}
\def\dij{\delta_{\sss ij}}
\def\gij{g_{\sss ij}}
\def\Gij{g^{\sss ij}}
\def\gkm{g_{\sss km}}
\def\Gkm{g^{\sss km}}
\def\cd#1{{}_{\sss;#1}}
\def\ud#1{{}_{\sss,#1}}
\def\udu#1{{}^{\sss,#1}}
\def\upcd#1{{}_{\sss;}{}^{\sss #1}}
\def\upud#1{{}_{\sss,}{}^{\sss #1}}
\def\ro{r\_{0}}
\def\vro{{\bf r}\_{0}}
\def\qo{q\_{1}}
\def\qt{q\_{2}}
\def\lfd{L\&{D}_f}
\def\rar{\rightarrow}
\def\deriv#1#2{{d#1\over d#2}}
\def\oot{{1\over 2}}
\def\pdline#1#2{\partial#1/\partial#2}
\def\pdd#1#2#3{{\partial^2#1\over\partial#2\partial#3}}
\def\av#1{\langle#1\rangle}
\def\avlar#1{\big\langle#1\big\rangle}
\def\div{{\vec\nabla}\cdot}
\def\grad{{\vec\nabla}}
\def\curl{{\vec\nabla}\times}
\def\DD{{\cal D}}
\def\FF{{\cal F}}
\def\LL{{\cal L}}
\def\MMs{{\cal M}}
\def\MM{\lrartop{\MMs}}
\def\NN{{\cal N}}
\def\AA{{\cal A}}
\def\BB{{\cal B}}
\def\SS{{\cal S}}
\def\PPs{{\cal P}}
\def\VV{{\cal V}}
\def\UU{{\vec{\bf\cal U}}}
\def\vd{\VV\_{D}}
\def\GG{{\cal G}}
\def\GGq{\GG_q}
\def\PP{\lrartop{\PPs}}
\def\EPS{{\cal E}}
\def\Eij{E_{ij}}
\def\Aij{\AA_{ij}}
\def\gf{\grad\f}
\def\rgf{{\bf r}\cdot\grad\f}
\def\bs{{\bf s}}
\def\bd{{\bf d}}
\def\be{{\bf e}}
\def\bF{{\bf F}}
\def\br{{\bf r}}
\def\bR{{\bf R}}
\def\bJ{{\bf J}}
\def\bV{{\bf V}}
\def\ba{{\bf a}}
\def\bA{{\bf A}}
\def\bg{{\bf g}}
\def\bn{{\bf n}}
\def\vh{{\bf h}}
\def\vu{{\bf u}}
\def\va{{\bf a}}
\def\vb{{\bf b}}
\def\dva{\d{\bf a}}
\def\vv{{\bf v}}
\def\ve{{\bf e}}
\def\vp{{\bf p}}
\def\vq{{\bf q}}
\def\vx{{\bf x}}
\def\vA{{\bf A}}
\def\bI{{\bf I}}
\def\vI{\bI}
\def\gfds{\grad\f\cdot \bd\bs}
\def\hm{\hat\mu}
\def\hFF{\hat\FF}
\def\eegg{\be\otimes\be\grad\grad}
\def\inv{\int\_{V}}
\def\invv{\int\_{v}}
\def\ins{\int\_{\Sigma}}
\def\inss{\int\_{\sigma}}
\def\vi{v_i}
\def\sf{S_f}
\def\si{S_i}
\def\ds{\bd\bs}
\def\dtr{d\&3r}
\def\dDr{~d\&{D}r}
\def\dDh{~d\&{D}h}
\def\dtv{d\&3v}
\def\f{\varphi}
\def\fb{\f\&{B}}
\def\eq{E_q}
\def\fl{\varphi\_{L}}
\def\gfl{\grad\fl}
\def\fx{\f\_{x}}
\def\b{\beta}
\def\z{\zeta}
\def\bo{\beta\_{0}}
\def\bx{\beta\_{x}}
\def\d{\delta}
\def\c{\gamma}
\def\co{\gamma_0}
\def\t{\tau}
\def\s{\sigma}
\def\r{\rho}
\def\rb{\r\&{B}}
\def\Fq{\vF_q}
\def\hr{\hat\r}
\def\l{\lambda}
\def\m{\mu}
\def\n{\nu}
\def\a{\alpha}
\def\abs#1{\vert #1\vert}
\def\abgf{\abs{\gf}}
\def\abgfl{\abs{\gfl}}
\def\eps{\epsilon}
\def\pd#1#2{{\partial#1 \over \partial#2}}
\def\ad{\a\_{D}}
\def\apj{Astrophys. J.}
\def\bk{\par\noindent}
\def\vr{\br}
\def\vR{\bR}
\def\vf{{\bf f}}
\def\vg{\bg}
\def\vF{\bF}
\def\vn{\bn}
\def\mi{m\_{i}}
\def\qi{q\_{i}}
\def\gfs{(\gf)\&{2}}
\def\hnmo{\hat\n^{-1}}
\def\El#1{E(#1\vr_1,...,#1\vr\_{N})}
\def\p{\partial}
\def\Fmn{F_{\m\n}}
\def\FMN{F^{\m\n}}
\def\Jm{J^{\m}}
\def\Am{A_{\m}}
\def\e{a_0}
\def\es{\e^2}
\def\hg{G}
\def\gz{\grad\z}
\def\fa{\f_{\a}}
\def\ra{\r_{\a}}
\def\rba{\rb_{\a}}
\def\ff#1#2{\f\ud{#1}\ud{#2}}

\title{ SUSPENSION AND LEVITATION IN NONLINEAR THEORIES}
\author{Mordehai Milgrom}
\address{Department of condensed-matter physics,
Weizmann Institute, Rehovot Israel}
\maketitle
\begin{abstract}
I investigate stable equilibria of bodies in potential fields satisfying
a generalized Poisson equation $\div[\m(\abgf/\e)\gf]\propto\r$. This
describes diverse systems such as nonlinear dielectrics, certain flow
problems, magnets, and superconductors in nonlinear magnetic media;
equilibria of forced soap films; and equilibria in certain nonlinear
field theories such as Born-Infeld electromagnetism. Earnshaw's theorem,
totally barring stable equilibria in the linear case, breaks down. While
it is still impossible to suspend a test, point charge or dipole,
one can suspend point bodies of finite charge,
or extended test-charge bodies. I examine circumstances under
which this can be done, using limits and special cases. I also consider
the analogue of magnetic trapping of neutral (dipolar) particles.
\vskip 9pt
PACS numbers:
\end{abstract}
\section{ Introduction}
 The possibility to suspend bodies, or to levitate them against the pull
of gravity, by applying various long-range fields to them, is of
obvious interest and importance.
 This has to be done by fields whose
sources are outside the body to be suspended.
The fields usually considered (electrostatic, magnetostatic,
nonrelativistic gravitational) are derived from scalar potentials
that satisfy the Laplace equation outside sources.
There is a sweeping  statement, under the name of Earnshaw's
 theorem \cite{ear},
 of the impossibility to suspend, statically,
 in stable equilibrium, a body carrying any combination of the different
charges (in rigid distributions)
 in any setting of the fields, when there is no
overlap between suspended charges and sources.
This has to do with the fact that a solution of the Laplace
equation can attain true extrema only on the boundary of the domain
of solution--the maximum (extremum) principle, see e.g.\cite{gt}.
 \par
Many of the Earnshaw-type results, beyond the basic statement
of the impossibility to suspend a point charge,
 rest heavily on the linearity of the Laplace operator.
The Earnshaw statement may thus be expected to
break down in non-linear systems.
Here I revisit the question of stable equilibria of bodies
subject to fields described by a nonlinear generalization of the
Poisson equation:
 A source distribution $\r(\vr)$, in $D$-dimensional Euclidean
space, produces a potential $\f$ through
\beq \NN\f\equiv\div[\m(\abgf/\e)\gf]=\ad G\r, \label{i} \eeq
derivable from the action
\beq S= -\int~\r\f\dDr
-{\e^2\over 2\ad G}\int\FF[\gfs/\e^2]\dDr.
\label{action} \eeq
Here,
 $\e$ is a constant with the dimensions of $\gf,$
$G$ is a coupling constant, $\ad=2(\pi)\&{D/2}/\Gamma(D/2)$
 is the $D$-dimensional solid angle (introduced for convenience),
$\m(x)=d\FF(y)/dy$ ($y=x^2$) is positive except perhaps that $\m(0)$ may
vanish. I normalize to $\FF(0)=0$.
The nonlinear operator $\NN$ generalizes the Laplacian, and can
also be written as
$\NN=\m\Aij\p_i\p_j$, with $\Aij=\d_{ij}+\hm\f\ud{i}\f\ud{j}
/\abs{\gf}^2$, where $\hm\equiv x\m'/\m$.
 Points where $\gf=0$ need special treatment
in theories where $\m(0)=0$,
but this does not modify any of the results, and is ignored for brevity.
(Summation over repeated indices is implied throughout.)
Only the case where eq.(\ref{i}) is elliptic is considered,
 so $\Aij$ is positive definite, tantamount to $\hm>-1$.
 For $G>0$, like, point sources attract each other, and opposite
sources repel (as in gravity), and {\it vice versa} for $G<0$
(as in electostatics)\cite{stat}.
\par
In \cite{stat}
I derive general results pertaining to forces on bodies in such
theories. Here I concentrate on the
existence of, and criteria for, stable equilibria of bodies.
Solving the general problem requires numerical means, but much can be
learnt from closed-form solutions that can be found in certain limits and
special cases.
\par
 Equation(\ref{i}) describes a variety of physical problems.
Some examples are:
(a) Nonlinear dielectric, and diamagnetic, media; $\m$ is then the
dielectric, or diamagnetic coefficient, which depends on the field
strength ($G<0$).
 (b) Subsonic-potential-flow problems of non-viscous fluids
with an equation of state of the form $p=p(\varrho)$ (subsonicity is
equivalent to ellipticity). The form of $\m(x)$
depends on the equation of state. For instance, when
 $p\propto \varrho^{\c}$,
 $\m(x)\propto [1-(\c-1)x^2]^{1/(\c-1)}$ for $\c\ge 1$ (the limit
 $\c\rar 1$ exists: ${\rm exp}(-x^2)$) ($G>0$). Our results then
 pertain to equilibria of sources, sinks, and obstacles in the flow.
(c) Equation(\ref{i}) was used in \cite{adl} as an effective-action
approximation to Abelianized QCD, with $\m(x)\propto lnx$; it is
 elliptic for $x<{\rm e}^{-1}$, or $x>1$.
(d) Nonlinear (vacuum) electrostatics as formulated e.g. in the
Born-Infeld nonlinear electromagnetism, which also appears in
 effective Lagrangians resulting from string theory
 (see review and references in \cite{gibra}\cite{gib}).
In the electrostatic case $\m(x)\propto (1-x^2)^{-1/2}$, and $G<0$.
(e) A formulation of an alternative nonrelativistic gravity to
replace the dark-matter hypothesis in galactic systems \cite{bm}.
Here $\m(x)\approx x$
 for $x\ll 1$, and $\m\approx 1$ for $x\gg 1$ ($G>0$).
(f) Problems of nonlinear electric-current flows in systems with
 field-dependent conductivity  (nonlinear
current-voltage relation), and nonlinear diffusion problems;
 $\m(x)$ is the transport coefficient. Here, a force on a body signifies
 the gradient of the entropy-production
rate with position of the body, so stable equilibria are configurations
of extremal entropy generation rates.
(g) Area (volume) minimization problems: If $\f(\vr)$ is understood as
the height of a $D$-dimensional surface above position $\vr$ on
 a $D$-dimensional hyperplane $H$, then eq.(\ref{i}) describes the
 problem of the minimization of the volume of the surface.
The sources could represent a prescribed vertical-force density.
 Forces on sources as will appear below correspond to
 lateral forces (parallel to $H$).
 In this problem $\m(x)\propto (1+x^2)^{-1/2}$, and $G>0$.

\section{The general problem}
\par
I start with the analogue of Earnshaw's original question:
can an equilibrium configuration $\r(\vr)$ exsit subject to
the $\f$ field alone? This would require that $\r\gf$ vanish everywhere,
and can probably be precluded, as I show for a wide class of theories.
In \cite{stat} I derive an expression for the virial integral
$\VV\equiv \int\r\vr\cdot\gf\dDr={\es/2\ad G}\int\FF(D-2\hFF)\dDr$
[with $\hFF(y)\equiv y\FF'(y)/\FF(y)$], that holds for theories in which
 $\hFF(0)<D/2$. (In this case the potential vanishes asymptotically
for a bounded charge; for $\hFF(0)=D/2$ it diverges logarithmically.)
We learn from this that in problems with $\hFF(y)< D/2$ we have
 $\VV\not =0$, and $\r\gf$ cannot vanish everywhere  (actually true
even for $\hFF=D/2$ where the expression for $\VV$ has an extra term).
This applies
 e.g., to the flow, and volume-minimization, problem (where $\hFF\le1$),
and to the modified dynamics (where $1\le\hFF\le D/2$).
Such complete-equilibrium configurations seem, anyway, to be of academic
 interest only. The bodies making up the systems we want to keep in
 equilibrium are themselves not held together, internally, in static
equilibrium by the same forces:
atoms and stars consist of moving constituents, held, intrinsically by
 forces other than electricity, and gravity, respectively.
\par
 Consider then equilibria of a charged body $B$,
given by a rigid charge distribution $\rb(\vr)$, in the
 presence of some fixed distribution $\hr(\vr)$
 of "holding" sources, with no overlap: $\hr\rb=0$. (Body and sources
might be each held rigid by forces other than the $\f$ field.)
The force on $B$ is
\beq  \vF=-\int\rb\gf\dDr,  \label{force} \eeq
with $\f$ determined from eq.(\ref{i}) with $\r(\vr)=\rb(\vr)+\hr(\vr)$.
$\vF$ is also the gradient of the energy with
 respect to rigid translations of $B$ by $\dva$:
$ \d E=-\dva\cdot\vF$ \cite{stat}.
I also want to consider the possibility of several
types of charge $\ra$ coupled to fields $\fa$ through
equations of type (\ref{i}), possibly with different forms of $\m(x)$
(e.g. levitation in some $\f$ field against gravity).
Under $\r\rar -\r$, $\f\rar -\f$ ,
 and $\vF\rar \vF$; under $G\rar -G$, $\f$ and all forces change sign.
\par
 Under what conditions
(choice of $\m(x)$, boundary conditions, $\hr$, and $\rb$) can $B$ be
suspended stably (with $E$ at a minimum)? I am mainly concerned
with translational stability, so define $E(\vR)$ as the translational
enery as a function of the position of some reference point
 $\vR$ in $B$, keeping the orientation fixed. $\vF=-\grad_{\vR}E(\vR)$.
A stable equilibrium where, $\vF=0$, requires that
 the Hessian of $E$, $\Eij=\pdd{E}{R_i}{R_j}$, be positive definite.
 (More generally, that
the first non-vanishing derivative-which must be of even order--is
positive definite: $E\ud{i_1}...\ud{i_{2n}}a_{i_{1}}...a_{i_{2n}}>0$
for any non-zero $\va$.)
\par
In the linear case we can separate $\f=\hat\f+\fb$, with $\hat\f$, and
$\fb$ coming, respectively from the "holding" sources, and from $B$
alone. The contribution of $\fb$ to the force in eq.(\ref{force})
 vanishes, because a body does not exert a force on itself, and we
then have (for the many-field case)
\beq E(\vR)=\sum_{\a}\int \rba(\vx)\hat\fa(\vR+\vx)dv
  \label{energy} \eeq
($\vx$ is the position inside the body). Clearly, when
all $\hat\fa$ satisfy the Laplace equation,
 so does $E(\vR)$, and $E$ does not have
a true extremum, leading to the general Earnshaw statement.
\par
In the nonlinear case eq.(\ref{energy}) remains valid only when
$B$ is a test body, whose contribution to the sources can be neglected
when calculating the potential. For a point, test body, in a single field
we have $E(\vR)=q\hat\f$, and
 it is still not possible to have a stable equilibrium because
 solutions $\hat\f$ of eq.(\ref{i}) still satisfy an extremum principle.
However, any departure from this restricted case might permit stable
 suspension.
(i) Unlike a test point charge, a point
 body of finite charge is not denied a
 stable equilibrium. The extremum principle for $\hat\f$ does
not exclude this, as $\hat\f$ is no more the effective
 potential (energy) of the body.
(ii) Even test charges arranged rigidly in an extended body can sometimes
 be suspended in a $\f$ field.
(iii) A point, test particle that carries test charges of different types
can be suspended in a combination of fields $\fa$
 satisfying eq.(\ref{i}). Now,
$E=\sum\_{\a}q\_{\a}\fa$ is not subject to an extremum
 principle even though the separate $\fa$s are.

\section{ Equilibria of point charges}
\par
 Take the case of a point charge $q$ "held" by a
 distribution $\hr$, when the only boundary
condition dictated is $\f\rar 0$ at infinity.
The problem can be solved in closed form
in the limit where $q$ is very large (relative to $\hr$).
 The total force on
the whole system vanishes, so the force on $q$ is equal and opposite
that on $\hr$. In the above limit, $\hr$ may be treated as a distribution
of test charges, so the force on it is
 $-\int\dDr \hr(\vr)\grad\fb$, where $\fb$ is
 produced by $B$ alone. This is gotten straightforwardly by applying
Gauss theorem to eq.(\ref{i}) for a point charge. Putting all this
together we obtain
for the force $\Fq(\vR)$ on $q$, at position $\vR$:
\beq \Fq(\vR)=s(q\hg)\e
\int\dDr\hr(\vr)\SS\left({\abs{q G}\over \e \abs{\vr-\vR}\&{D-1}}
\right){\vr-\vR\over\abs{\vr-\vR}}. \label{lutaru} \eeq
Here, $s(x)=sign(x)$, and
 $\SS(y)$ is the inverse of $\n(x)=x\m(x)$ ($\n$ is increasing, by
the ellipticity condition).
The energy $E$ ($\Fq(\vR)=-\grad\_{\vR}E$) is thus linear in $\hr$:
\beq E(\vR)=\int\dDr\hr(\vr)\GGq(\abs{\vr-\vR}).
 \label{brazar} \eeq
>From eq.(\ref{lutaru}),
the effective Green's function, $\GGq$, is integrated from
  ($x\not =0$ in light of the non-overlap assumption)
\beq \grad_{\vx}\GGq(\vx)=s(q\hg)\e
\SS(z){\vx\over \abs{\vx}},  \label{gutareq} \eeq
with $z=\abs{q G}/ \e \abs{\vx}\&{D-1}$.
The Hessian of $\GGq$ is
\beq \GGq\ud{ij}=s(q\hg)\e{\SS(z)\over\abs{\vx}}\BB_{ij},~~~
\BB_{ij}=\d_{ij}-{D+\hm\over 1+\hm}{\ve}_i{\ve}_j,
 \label{payeru} \eeq
with ${\ve}_i=x_i/\abs{\vx}$. $\BB$ has $D-1$ eigenvalues equal $1$,
and one that equals $-(D-1)/(1+\hm)<0$. So,
\beq \Delta \GGq(\vx)=s(q\hg)(D-1)\e
\SS(z)\abs{\vx}^{-1}{\hm(z)\over 1+\hm(z)}  \label{qurates} \eeq
($\hm(z)=z\m'/\m$). Also,
\beq \Eij(\vR)=\int\dDr\hr(\vr)\GGq\ud{ij}(\abs{\vr-\vR}).
 \label{bhoutr} \eeq
Since $\SS$ depends nontrivially on its argument,
the dependence of $\Fq$, $\GGq$, and $E$ on $q$ can be
 nontrivial. Among other things, the location of equilibria may depend
 on $q$.
Take $\hr$ to have a center of symmetry, which must then be an
equilibrium point; when is it stable?
If $\hr$ has cubic symmetry the question hinges on the
sign $\Delta E$ at the center (taken at the origin),
because from the symmetry
\beq \Eij(0)=D^{-1}\Delta E(0)\d_{ij}.
\label{patsa}\eeq
If $\hm$ and $\hr$ have fixed signs, then from
 eqs.(\ref{brazar})(\ref{qurates})
 $\Delta E$ has the fixed sign $s=sign(q\hm G\hr)$
at all $\vR$ for which there is no overlap. If $s$ is positive,
$\Delta E>0$, and there might be a true
 minimum (while $\Delta E<0$ precludes a stable equilibrium).
 In particular, in the
cubic-symmetric case the center is then per force a stable equilibrium.
By the same token, the charge distribution $\hr$ can
 be suspended (stably against translations) in the field of $q$,
 exemplifying statement (ii) below eq.(\ref{energy}).
Thus, e.g., in the flow, Born-Infeld, and volume-minimization problems,
 where $\hm G<0$, $\hr$ has to
have a sign opposite to that of $q$ to be able to suspend $q$.
If the cubic symmetry of $\hr$ is slightly disturbed there
 remains an equilibrium point (from continuity), which then may depend
on $q$ if the center of symmetry is lost.
\par
When $\hr$ has lower-than-cubic symmetry, the center might still be a
stable equilibrium: if $\GGq\ud{ij}$ does not have a
definite sign, $\Eij$ might be positive definite upon
integration by eq.(\ref{bhoutr}). For
 example, in $D>2$, if $\hr$ is any planar (two-dimensional) distribution
with a four-fold symmetry,
 angular integration of $\BB$ gives a matrix whose
eigenvalues are all $1$, but two that equal $(\hm-D+2)/2(1+\hm)$.
We get a stable minimum at the center
 if $\hm(z)>D-2$, for all $z$ contributing to the radial integration.
This is straightforwardly generalized to other symmetries.
Clearly, because of the linearity in $\hr$ stability of a composite
configuration can be assessed by adding up $\Eij$ of the components.
So, any combination of configurations that each
gives a stable equilibrium also does so (e.g. a combination
of any number of concentric, cubic-symmetric $\hr(\vr)$).
\par
What happens when $q$ is decreased away from the large-$q$ limit? In the
 general case, equilibrium might conceivably be lost,
as $q$ becomes comparable with the the "holding" charges, when
one of the eigenvalues of $\Eij$ changes sign.
For cubic-symmetric $\hr$, however,
 where eq.(\ref{patsa}) is valid for all $q$, the answer depends on the
sign of $\Delta E(0)$. In the test-particle limit,
 $q\rar 0$, we have $E\rar q\hat\f$. Since the Hessian
of $\hat\f$ cannot have a definite sign at the center, we must have
 $d_q\equiv q^{-1}\Delta E(0)\rar \Delta\hat\f(0)\rar 0$ in the limit.
It is likely that stability is lost in the test-charge limit
with $d_q$ remaining positive all the way down to $q=0$. Then, the
center remains a stable equilibrium for all finite $q$. If $d_q$ changes
sign and approach 0 from below, then the same configuration has a stable
equilibrium in a theory with $G\rar -G$. (The possibility that
$d_q=0$ for a finite strech of $q$ may be rejected from analyticity. In
 the linear case $d_q=0$ for all $q$.)
\par
The above treatment (for large $q/\hr$) is applicable
 whenever we have the solution for the field
$\fb$ of $B$ alone. For instance, all the above expressions still
hold if $B$ is any spherically symmetric charge
 (not overlapping with $\hr$), with $q$ its total charge. Exact solutions
can also be found for any body with 1-D symmetry (plane, cylindrical,
 etc.). Another example:
 in \cite{conf} I derive the exact field for a pair of opposite point
 charges
in the theory with $\m(x)\propto x^{D-2}$; so its equilibria can also
be studied in the limit of weak "holding" sources.

\section{ Some impossibilities}
\par
Even if stable equilibria are possible in the nonlinear case, some
elementary feats are still impossible (a)
The impossibility to suspend a point, test charge in a single
field stems from the fact that
 $E=q\hat\f$, and so $\Aij\Eij=0$ away from sources. As $\AA$
 is positive definite, $\Eij$ cannot be positive definite as stability
requires. (I assume here and below that not all the $\Eij$ vanish
at the equilibrium point. More generally, it can be shown
that the first, relevant (even) derivative that does not vanish
cannot be positive definite.)
 (b) It is impossible to balance
stably a test point charge by two $\f$ fields in a region where one of
them has a constant gradient $\gf'=\va$ (e.g. levitate the charge in a
 $\f$ field against earth-surface gravity). Here
 $E=q(\va\cdot\vr+\f)$, and also satisfies
$\Aij(\gf)\Eij=0$ away from sources, and so $E$ cannot have
a minimum.
 (c) Even if extended, test bodies can, in general be suspended,
 it is not possible to suspend a test dipole in a $\f$ field.
 The force on a dipole $\vp$ is $-(\vp\cdot\grad)\gf$. The
translational energy is then $E=\vp\cdot\gf$, and
$\Aij(\gf)\Eij=(\vp\cdot\grad)(\Aij\f\ud{i}\ud{j})
-\f\ud{i}\ud{j}\pd{\Aij}{\f\ud{k}}(\vp\cdot\grad)\f\ud{k}$.
The first term vanishes everywhere away from sources; the second term
vanishes at equilibria, because $(\vp\cdot\grad)\gf$ is the
 force. Again, $\Eij$ cannot be positive
definite at an equilibrium--no stable minimum, not
 even against translations.
 (d) Consider now a small
 test body that has both a charge $q$ and a dipole $\vp$ in equilibrium
at a point where $\gf\not = 0$. Here $E=q\f+\vp\cdot\gf$.
As in case (c) we have
$\Aij\Eij=-\f\ud{i}\ud{j}\pd{\Aij}{\f\ud{k}}
(\vp\cdot\grad)\f\ud{k}$,
but now the vanishing of the force implies
 $(\vp\cdot\grad)\f\ud{k}=-q\f\ud{k}$, so
$\Aij\Eij=q\f\ud{i}\ud{j}\pd{\Aij}{\f\ud{k}}
\f\ud{k}$.
Take, specifically, the orientation for which the moment on the
dipole vanishes (also a requirement of equilibrium): $\vp=-p\gf/\abs
{\gf}$ ($p>0$)
 calculated at the equilibrium point, and then consider stability
against pure translations. For this value of $\vp$ one finds
 that at the equilibrium point
$\Aij\Eij=p^{-1}q^2\abs{\gf}^2\deriv{\hm(x)}{x}$. (putting $\e=1$.)
A stable equilibrium might then be possible only where
$\hm$ is increasing. This is precluded, e.g., in the flow,
and the volume-minimizaton, problem, and in the modified
dynamics, where $\hm$ is always decreasing,
 but not in Born-Infeld electrostatics. As we have only
checked stability to translations, increasing $\hm$ does not
guarantee full stability.

\section{ Suspension of aligned dipoles}
\par
Neutral, dipolar bodies can be levitated
stably, even in the linear case, provided the (constant-magnitude)
 dipole is forced to remain
aligned with $\gf$. For example, under conditions of adiabaticity,
 precession of a spin (aligned with the dipole) about $\gf$ insures
alignment. This fact is used,
e.g., in the construction of magnetic traps for atoms and neutrons
\cite{paul}, and for macroscopic, magnetized tops\cite{ber}.
Take then $\vp=-p\ve$, where $\ve=\gf/\abs{\gf}$
(restricted to equilibria at points where $\gf\not = 0$), and
 consider levitation of the body against a constant-force field $\vf$.
The total force is $\vf+p\grad\abs{\gf}$,
the energy (for translation plus realignment) is
 $E(\vR)=-\vf\cdot\vR-p\abs{\gf}$,
and $\Eij=-p\abs{\gf}^{-1}(\ff{k}{i}\ff{k}{j}
-\ff{z}{i}\ff{z}{j}+\abs{\gf}\ff{z}{i}\ud{j})$, with $z$ the
 direction of $\ve$ (and is not summed over).
 As before, we require, as a necessary condition for
stability,  $\Aij\Eij>0$ at the equilibrium, where we can put
$\ff{z}{i}=-p^{-1}f_{i}$. After some algebra we get
\beq \Aij\Eij  =
-p^{-1}\abs{\gf}^{-1}[p^2\ff{i}{k}\ff{i}{k}-\abs{\vf}^2
\nonumber\\
~~~ +  \hm(f_z^2-\abs{\vf}^2)-\hm'\abs{\gf}f_z^2].  \label{potar} \eeq
In the linear case ($\hm=0$) the expression in parentheses is positive
(unless all the $\ff{i}{j}=0$.) and we restore $p<0$ as the necessary
condition. In the case of suspension in a pure $\f$ field ($\vf={\bf 0}$)
it remains so in the nonlinear case. When $\vf\not ={\bf 0}$,
 stability depends also on $\hm$ and $\hm'$.
 Take, e.g., the case where $\vf$ is along a symmetry axis of the field
so $f_z^2=\abs{\vf}^2$. Then,
$\Aij\Eij=-p^{-1}\abs{\gf}^{-1}[p^2\ff{i}{k}\ff{i}{k}-\abs{\vf}^2
(1+\hm'\abs{\gf})]$. With large enough $\hm'$ it may be possible
to get stable suspension even with $p>0$. For example, when can such a
dipole be levitated in the $\f$
 field of a point charge against a constant $\vf$? I find that this can
be done if and only if one takes $p>0$ (dipole aligned with $-\gf$),
and $\hm$ satisfies at the equilibrium position
 $\hm\deriv{{\bf ln}\hm}{{\bf ln}x}>1+(1+\hm)/(D-1)$.
Of the problems listed above, the condition on $\hm$ can be met only
in the Born-Infeld theory; in the others $\hm\hm'<0$.

\end{document}